# Artificial Intelligence in a Photonic Temporal Processor

Youlve Chen[1]†, Jinlong Xiang[1]†, Yimin Hu[1]†, Yuchen Yin[1], Chaojun Xu[1], Zhengshun Lei[1], Xin Wang[1], Yufeng Zhang[2*], Yixiao Zhu[1,*], Qunbi Zhuge[1], Junwen Zhang[3], Wei Chu[4],Tao Lin[5], Yikai Su[1], Zhipei Sun[2,*], Xuhan Guo[1,*]

[1.] School of Information Science and Electrical Engineering, State Key Laboratory of Photonics and Communications, Shanghai Jiao Tong University, Shanghai, China.

[2.] Department of Electronics and Nanoengineering, Aalto University, Espoo, Finland.

[3.] College of Future Information Technology, Fudan University, Shanghai, China

[4.] Zhangjiang Laboratory, Shanghai 201210, China

[5.] North Ocean Photonics Co. Ltd.,Shanghai, 201306, China.

Corresponding authors: Yufeng Zhang (yufeng.zhang@aalto.fi), Yixiao Zhu (yixiaozhu@sjtu.edu.cn), Zhipei Sun (zhipei.sun@aalto.fi), Xuhan Guo (guoxuhan@sjtu.edu.cn)

†These authors contributed equally to this work.

## Abstract:

Optical neural networks (ONNs) promise high-throughput and energy-efficient artificial intelligence, yet essentially all implementations so far encode information across space either in free-space arrays or in integrated waveguide meshes, tying the number of neurons to the number of physical components and fixes the routing topology at fabrication. Here we show that moving the computation into time decouples computational dimension from hardware dimension. Exploiting space-time duality, we implement optical diffraction and interference entirely in time domain, using thin-film lithium niobate modulators as time lenses and temporal masks, with chromatic dispersion providing the coupling between successive temporal neurons. We experimentally verify high-order, complex-valued matrix-matrix multiplications using just a single optical input/output port, scaling the computational dimensions far beyond the channel count. By incorporating optical feedback, we extend this platform into versatile neural networks, where the network layers, neuron numbers, and synaptic connections are fully programmable and *in-situ* trainable. Our temporal diffractive neural networks are successfully validated on various classification benchmarks, alongside image and video generation tasks. Notably, using this platform we demonstrate an all-analogue generative pipeline in which the latent variable is drawn directly from amplified spontaneous emission, so that no digital sampling or electronic modulation appears anywhere in the generative path. Furthermore, high-resolution images and videos are generated at high frame rates, outperforming state-of-the-art modulator-refresh-limited optical generative systems. These results establish a unified photonic temporal computing framework that encompasses foundational operators, adaptive architectures, and learning capacities, providing a scalable and deployable pathway toward next-generation machine intelligence.

The rapid expansion of generative artificial intelligence (AI) is placing unprecedented demands on computational power and energy consumption[1]. Traditional electronic infrastructure, increasingly constrained by von Neumann bottlenecks and the physical limits of transistor scaling, struggles to sustain these growing computational demands[2]. Optical neural networks (ONNs) have emerged as a compelling alternative by exploiting the wave nature of light so that optical propagation itself performs massively parallel multiply–accumulate (MAC) operations, offering substantial improvements in energy efficiency and processing latency[3-6]. To date, ONNs have been predominantly implemented in space[7]: as light propagates through either free space or an integrated waveguide circuit, diffraction and interference distribute signals across dense spatial connections. These principles have established two dominant architectures of ONN systems: diffractive networks[8-13] and matrix accelerators[14-25]. Enhanced by wavelength multiplexing and multilayer cascades, such platforms can execute the core linear primitives of modern deep learning, including convolution, full connection, and self-attention, supporting a diverse range of inference and generative models directly in the optical domain[26-37].

Spatial ONNs, however, remain fundamentally constrained by physical and system-level bottlenecks. Scaling such schemes requires photonic component arrays that grow quadratically with network dimensions, demanding stringent channel uniformity and independent control of many spatial degrees of freedom. This necessitates extensive peripheral electronics and complex calibration and control protocols, which in turn introduce latency, crosstalk, and instability—issues prevalent in spatial-light-modulation systems, optical sensor arrays, and photonic integrated circuits (PICs). Moreover, the physical dimensions of these spatial field-manipulation elements are ultimately bounded by the optical wavelength, preventing optical spatial platforms from approaching the integration density of advanced electronics. This scalability bottleneck is further compounded by a persistent trade-off in optoelectronic input/output (I/O) interfaces: free-space systems accommodate massive parallel channels but have restricted single-channel modulation bandwidth, whereas PICs deliver ultra-high per-channel data rates but suffer in spatial multiplexing scale and neuron capacity. Beyond hardware and I/O constraints, spatial platforms fail to address the structural diversity of modern AI. Since the routing of spatial ONNs is set at fabrication, neuron counts and network topology remain inherently static and cannot dynamically adapt to the architectural diversity of contemporary machine learning.

Temporal optical computing removes these barriers by shifting the computing dimension from space to time. Space–time duality establishes that paraxial diffraction is mathematically identical to narrowband dispersive propagation, and that a quadratic phase in space (a lens) has an exact counterpart in a quadratic phase in time (a time lens)[38-40]. Built on this exact correspondence, loop-based temporal optical neural networks in which a single optical unit performs every multiply–accumulate operation and successive round trips realise successive layers have advanced rapidly: serial neurons coupled by group-delay dispersion[49], time-lens-based learning[50], recirculating depth via a temporal synthetic dimension[47], coupled fibre loops with programmable gain and in-loop training[42]. Here we implement a photonic temporal computing framework utilizing synchronized thin-film lithium niobate (TFLN) IQ modulators coupled with an optical dispersive medium. By configuring the high-speed TFLN modulator as a time lens and adopting a time-frequency encoding scheme, we achieve high-order, complex-valued matrix-matrix multiplications through a single optical I/O port, scaling the computational dimensions far beyond the physical channel count of the underlying hardware.

**Table 1. Comparison with SOTA photonic neural networks**

| | Source | Hardware | Architecture tunability* | Programmable photonic neuron | Input features | In-situ training | Throughput (TOPS) | Energy efficiency (TOPS/W) |
|---|---|---|---|---|---|---|---|---|
| Spatial ONNs | Nat. Photonics (2021)[41] | DMD + SLM + sCMOS | No | 1,470,000 (SLM) | 700×700 | Yes | 240.1 | 1.58 |
| | Nature (2023)[28] | SiO2 mask + EAC chip | No | 0 | 400×400 | Yes | $4.55\times10^{3}$ | $7.48\times10^{4}$ |
| | Science (2025)[33] | SiO2 phase masks + OLS | No | None | 1920× 1080 | No | $3.57\times10^{4}$ | 664 |
| | Nat. Photon. (2025)[13] | SOI chip | No | 64 (MRM) | 64 | Yes | $6.71\times10^{-5}$ | — |
| | Nature (2022)[26] | SOI chip | No | 66 (PIN Att.) | 30 | No | 0.27 | 0.07 |
| | Nat. Photon. (2024)[29] | SOI chip | No | 132 (MZI mesh) | 6 | Yes | 0.59 | 0.11 |
| | Science (2024)[30] | SOI chip | No | 160 (MZI mesh) | 64 | No | $5.04\times10^{4}$ | 160.82 |
| | eLight (2025)[32] | SOI chip | No | 12 (MZI mesh) | 2 | Yes | 0.0587 | 0.48 |
| Temporal ONNs | Nature (2021)[15] | Microcomb + EOM + Fiber | Yes | 90 | 250,000 | No | 11.3 | — |
| | Nat. Commun. (2026)[42] | PM + MZM + Fiber loop | Yes | 31,124 | 251 | Yes | $30.55\times10^{-6}$ | — |
| | **TDNN (This work*)** | **TFLN chip + Fiber** | **Yes** | **4000 (Temporal mask)** | **~1000 (Exp.)** | **Yes** | **500** | **266.1** |

*Here, architectural tunability refers to the system's capability to support diverse physical implementations, mathematical operations, and connectivity topologies characteristic of various neural network paradigms (e.g., fully connected, diffractive, convolutional, and Fourier neural networks).

We further introduce iterative temporal dynamics by incorporating an optical feedback loop, extending the same hardware to a multilayer temporal diffractive neural network (TDNN) without additional spatial resources or parallel optical paths. Crucially, network layers, neuron counts, and synaptic connections are fully programmable and *in-situ* trainable. The flexible operators and routing of temporal neurons support versatile architecture of neural networks. We validate the computing capabilities of TDNNs across both discriminative and generative machine learning tasks. We first demonstrate an all-analog, end-to-end generative pipeline that synthesizes random images directly from temporal Gaussian white noise. Furthermore, we generate high-resolution images and videos with ultra-fast frame rate (e.g., > 40 MHz for the resolution of 50×50), representing a higher speed improvement compared to state-of-the-art (SOTA) modulator-refresh-limited optical generative systems[33,34]. Table 1 compares our approach with SOTA optical neural networks (ONNs). Existing spatial architectures generally face trade-offs among network scale, programmability, throughput and hardware complexity, as increasing the number of neurons or input dimensions typically requires additional physical components and interfaces. In contrast, TDNN decouples the logical network architecture from the physical hardware by mapping neurons, connectivity and layer operations into the temporal domain. This enables diverse neural-network paradigms with different mathematical operators and connectivity topologies to be implemented within the same optical pathway. Experimentally, our system supports approximately 4,000 programmable photonic neurons and input dimensions approaching 1,000, while achieving a throughput of 500 TOPS and an energy efficiency of 266.1 TOPS/W. More importantly, the entire multilayer network is accessed through a single optical I/O interface with minimal electrical connections, allowing computational scale and throughput to increase without a corresponding growth in hardware and I/O complexity. Our results establish a comprehensive photonic temporal intelligence paradigm that retains the intrinsic high parallelism, dense connectivity, and ultrafast propagation of optics, while addressing the critical bottlenecks of spatial scalability and architectural reconfigurability. Its compatibility with optical fiber links and on-chip optical interconnects points toward a practical deployment pathway for next-generation photonic machine intelligence.

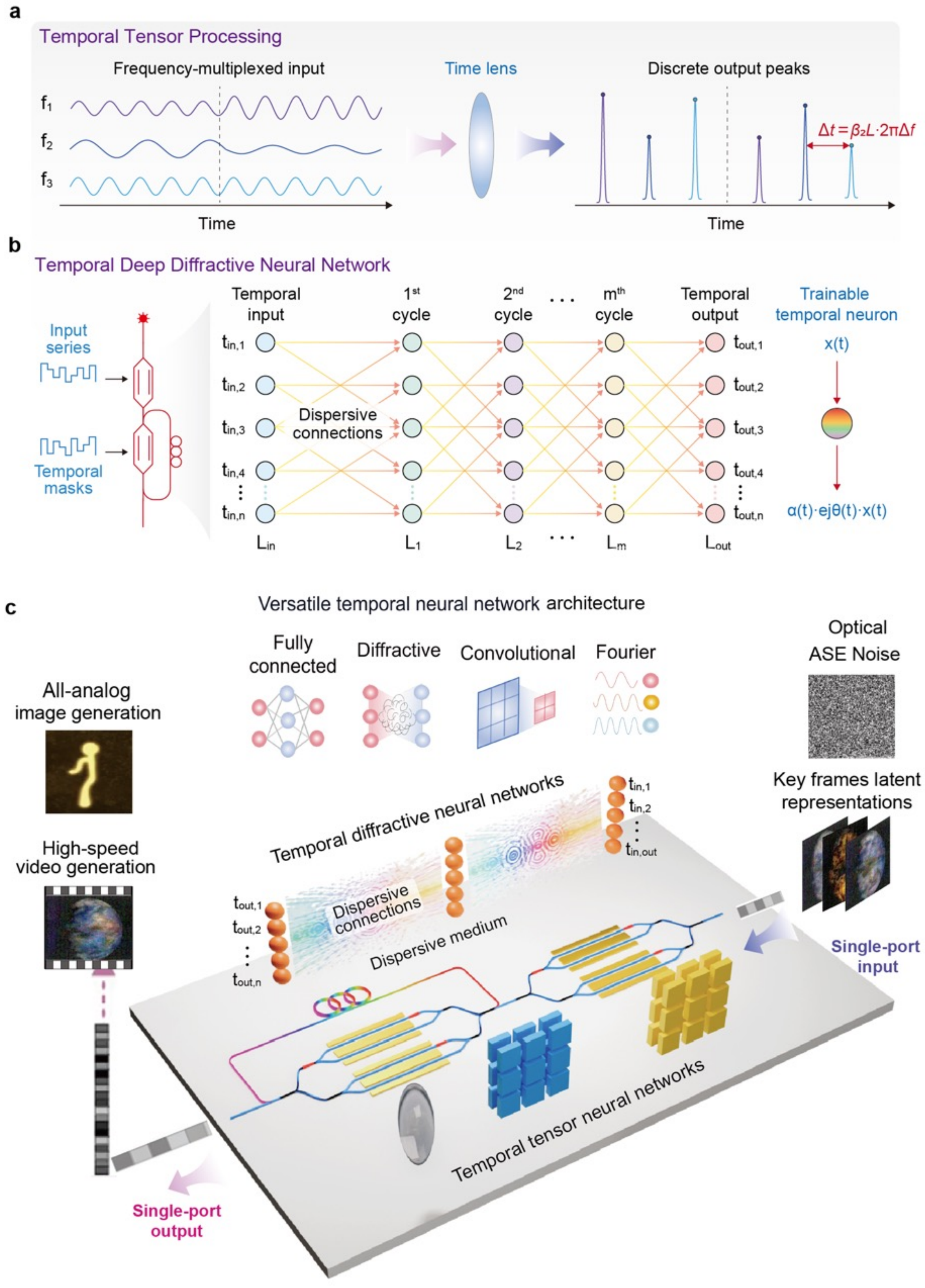

**Fig. 1 Concept of photonic temporal intelligence. a–b,** Fundamental computational primitives of photonic temporal computing. **a,** Time lens. A temporal optical waveform undergoing quadratic phase modulation followed by dispersive propagation is Fourier transformed in the temporal domain. This real-time optical Fourier transform enables tensor multiplication through time–

frequency encoding of input matrices. **b,** Temporal diffractive unit and multilayer temporal diffractive neural networks (TDNNs). Temporal neurons are represented by individual sampling points of the optical waveform and are independently modulated by programmable temporal masks with complex-valued weights. During dispersive propagation, temporal neurons coherently interact through the dispersive transfer function, analogous to spatial diffraction in diffractive neural networks (Supplementary Fig. 1b). Optical feedback enables multilayer TDNN architectures without requiring repeated analogue-to-digital or digital-to-analogue conversion of temporal representations. **c,** Implementation of the photonic temporal computing system. IQ modulators provide programmable control of multiple optical functions, including time-lens operation, complex-valued temporal masking and time–frequency tensor encoding. Dispersive media establish interactions between temporal neurons, while feedback loops enable recurrent propagation through multiple computational layers. By combining programmable temporal modulation with reconfigurable temporal connectivity, the TDNN supports diverse optical neural architectures, including fully connected, diffractive, convolutional and Fourier neural networks.

## Principle of photonic temporal intelligence

Figure 1a–b illustrates the fundamental computational primitives underlying photonic temporal intelligence, which originate from the space–time duality of optical propagation (Supplementary Fig. 1). The first primitive is the time lens. As shown in Fig. 1a, a temporal optical waveform undergoing quadratic phase modulation followed by dispersive propagation experiences a time-domain Fourier transformation. Consequently, the output waveform corresponds to the Fourier transform of the input signal, accompanied by a quadratic phase factor and temporal scaling (Supplementary Note 1). This real-time optical Fourier transformation establishes a fundamental mechanism for one-shot matrix multiplication through time–frequency encoding. Unlike conventional radio frequency multiplexing approaches that require frequency-resolved detection and subsequent electronic reconstruction, the calculated results here can be directly sampled in the temporal domain.

To realize a high-performance temporal lens, we employ an ultra-broadband IQ modulator (~60 GHz bandwidth) to generate broadband quadratic phase modulation with an extended temporal aperture of ~15 ns. The resulting Fourier transform achieves a frequency resolution of approximately 60 MHz. Compared with conventional time lens implementations based on nonlinear four-wave mixing, phase modulation or other nonlinear optical processes[38-40], this approach provides a programmable, stable and broadband platform for high-fidelity temporal imaging. Such advancements in temporal aperture and spectral resolution enables large-scale tensor multiplication within a compact optical architecture.

The second computational primitive is the temporal diffractive unit (Fig. 1b). Here, the IQ modulator functions as programmable temporal masks that assign trainable complex-valued weights to individual temporal neurons. A temporal neuron is defined as a discrete sampling point of the optical waveform, whose representation can be adapted according to the target computational task. Upon propagation through a dispersive medium, these temporal neurons interact with each other through the dispersive transfer function, producing coherent temporal-field evolution analogous to spatial diffraction in diffractive neural networks. By incorporating an optical feedback loop, the same physical optical path can be reused for multi-layer neural computation without intermediate analogue-to-digital or digital-to-analogue conversion. Each round trip corresponds to one computational layer of the temporal diffractive neural network (TDNN). Within each layer, the temporal neurons are transformed by a programmable temporal mask and subsequently propagated through dispersion, after which the resulting waveform is returned to the modulator for the next layer transformation. This temporal recurrence enables scalable network depth without increasing the physical footprint of the optical processor.

Figure 1c illustrates the overall TDNN architecture. By mapping the neural network representation into the temporal domain, large-scale optical neural networks can be implemented using a single optical input/output pathway with minimal electrical and optical interfaces. The high-speed programmability of IQ modulators allows the optical operator of each layer to be dynamically configured and trained. For example, tensor multiplication is realized by loading time–frequency encoded operands and temporal lens functions onto separate IQ modulators, whereas temporal diffraction is achieved by encoding input temporal neurons and layer-specific complex-valued masks onto programmable temporal modulators. Combined with dispersive propagation and optical feedback, these building blocks establish a reconfigurable forward-propagating TDNN capable of supporting diverse optical neural architectures, including fully connected, diffractive and Fourier neural networks, with convolutional processing implemented through the same time–frequency mapping (Supplementary Fig. 12), for a broad range of inference and generative tasks.

## Temporal tensor processing

Temporal neural networks naturally support tensor multiplication, a fundamental operation underlying modern artificial neural networks (ANNs). Figure 2a illustrates the implementation of general matrix–matrix multiplication (GEMM) using time–frequency encoding. The row and column dimensions of the input matrices are mapped onto temporal and frequency degrees of freedom, respectively. Specifically, matrix elements along each row are serialized into temporal waveforms, while different row vectors are encoded onto distinct radio-frequency carriers. For the first matrix, we define the spacing between adjacent row vector carrier frequencies as $\Delta f$, and the symbol duration of each row element as $1/\Delta f$. For the second matrix, the symbol duration of each row element is also $1/\Delta f$, but the the frequency spacing $\Delta F$ should cover the bandwidth of the first matrix (e.g., $\Delta F = n\Delta f$, $n$ is the number of row vectors in the first matrix). Two matrices are independently loaded onto separate IQ modulators, where element-wise multiplication is achieved through controlled temporal alignment between the encoded waveforms. One IQ modulator simultaneously performs quadratic temporal phase modulation to realize a time-lens operation, enabling real-time optical Fourier transformation after dispersive propagation.

The resulting Fourier transformation coherently accumulates the dot products between row-vector pairs, with the computational results encoded at frequencies corresponding to the sums of the input carrier frequencies. Therefore, the output matrix elements can be directly sampled in the temporal domain at the receiver (Fig. 2c), without requiring digital inverse Fourier transformation as used in conventional RF multiplexing approaches[17]. The temporal sampling interval $\Delta t$ is determined by the dispersion relationship: $\Delta t = \beta_2 L \cdot 2\pi\Delta f$, where $\beta_2$ denotes the group velocity dispersion (GVD) coefficient and $L$ is the dispersive propagation length. Higher-order tensor dimensions can be further introduced through an expanded frequency aperture or additional wavelength channels, providing a scalable route towards multi-dimensional tensor processing. Beyond matrix multiplication, the same framework can also support other tensor operations, including image convolution (Supplementary Fig. 12).

For real-valued positive tensors, the computed data streams can be directly obtained using an intensity direct-detection receiver. While for general complex-valued tensor operations, a coherent receiver is employed to retrieve both amplitude and phase information at each temporal sampling position (Methods and Supplementary Fig. 4). Figures 2b–c show experimental demonstrations of a 5×5 positive-valued MMM, where 25 output elements are directly sampled from a digital storage oscilloscope (DSO) at the expected temporal positions with low reconstruction error. Furthermore, Fig. 2b illustrates the wavelength consistency by presenting results at four different wavelengths, all of which show minimal errors. Figure 2d illustrates complex-valued matrix–vector multiplication (MVM), where the measured results closely agree with the target values.

Enabled by the large temporal aperture provided by the 60-GHz-bandwidth IQ modulator, our system achieves matrix dimensions of up to 35×35 for MVM and 11×11 for MMM. This is very close to the theoretical limit of the size achievable under the time-aperture bandwidth product (Supplementary Note 3). The detailed time–frequency encoding scheme is described in Methods. As shown in Fig. 2e, the overall computational precision exceeds 4 bits for MVM across all demonstrated sizes and approaches 4 bits for the largest MMM, where the precision is quantified as: $\log_2 \frac{(\mathrm{r_{max}} - \mathrm{r_{min}})}{\mathrm{RMSE_r}}$, with RMSE representing the root mean square error between the measured and target matrix elements, and $r_{max}$ and $r_{min}$ denoting the maximum and minimum target values, respectively. By exploiting four-wavelength parallelism (Fig. 2b and Supplementary Fig. 5), the achieved throughput for general-element MMM reaches 3.036 TOPS (Methods). Multi-wavelength parallelism provides a pathway for high-order tensor processing. Beyond simple parallel processing, synthetic temporal aperture offers a approach to performing summations in higher dimensions (Supplementary Fig. 15).

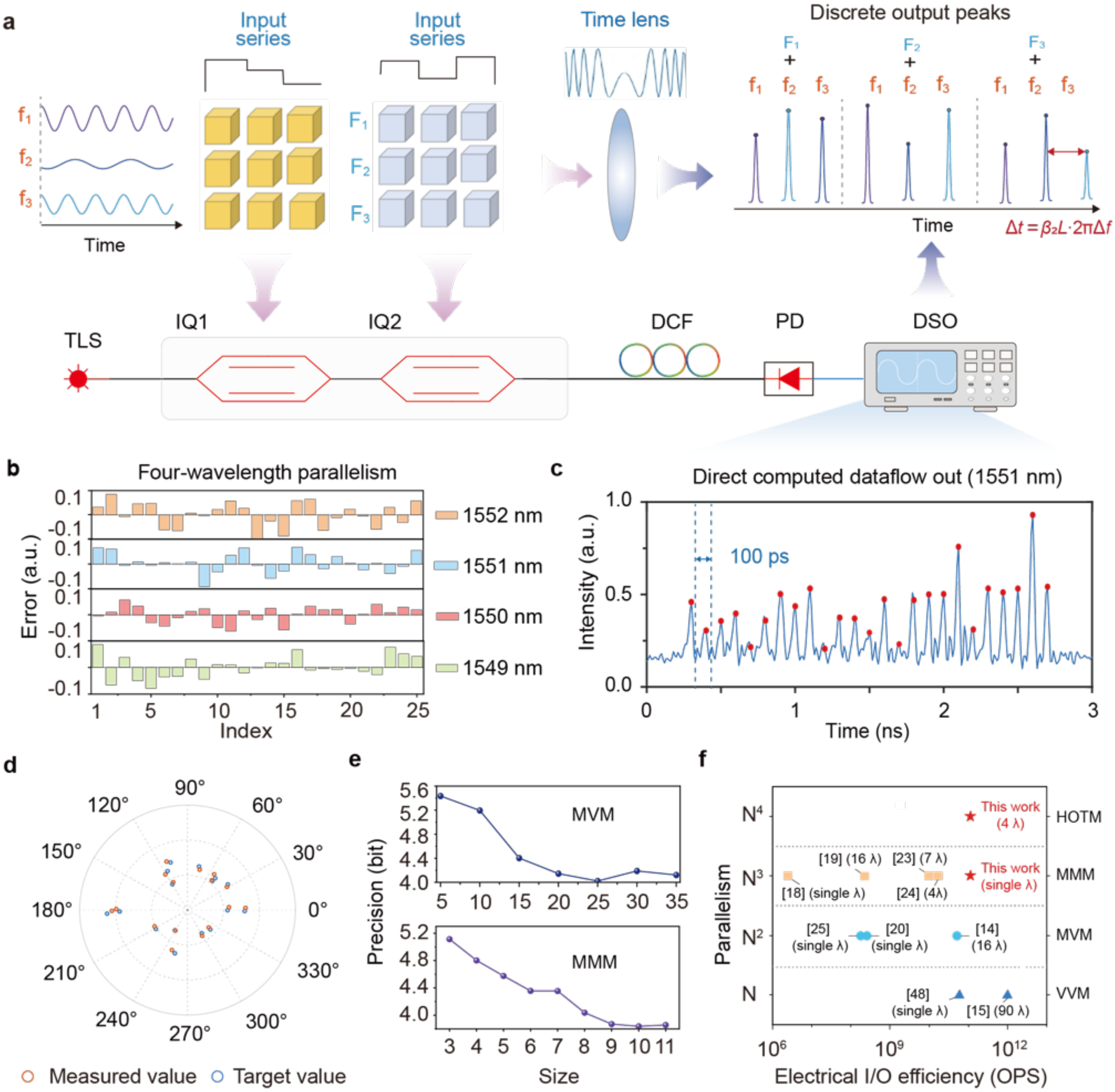

**Fig. 2. Photonic temporal tensor processing via time lens. a,** Principle of tensor multiplication. Matrices are time-frequency encoded by mapping row elements to sequential time series, where each row is modulated at a unique frequency. The two matrices, as well as a time lens (TL) are assigned to separate IQ modulators respectively. **b-c,** An example of experimental 5×5

matrix-matrix multiplication (MMM). **b,** Comparison of computing errors for all 25 elements between target and measured values. The results at four different wavelengths exhibit minimal errors, demonstrating excellent wavelength consistency. **c,** Acquisition of the 25 computed elements at regular intervals at the receiver. **d,** Experimental results of complex-valued matrix-vector multiplication (MVM), showing that the measured values agree with the target values. **e,** The computing precision of matrix-vector multiplication (MVM) and matrix-matrix multiplication (MMM) with different sizes. Up to $35 \times 35$ MVM and $11 \times 11$ MMM are demonstrated. **f,** Comparison with SOTA optical matrix accelerators, where the vertical axis represents the computational parallelism ($N^i$ denote that the time complexity required for a digital platform to carry out this computation is $O(N^i)$[18]). For the horizontal axis, OPs denotes operations per second. HOTM, high-order tensor multiplication. More details can be found in Supplementary Table 1.

Conventional photonic tensor accelerators employ various multiplexing schemes to increase computational throughput. However, scaling these architectures often requires a proportional increase in optical–electrical I/O interfaces for data encoding and readout, resulting in substantial packaging complexity and becoming a potential bottleneck for practical deployment. To quantify this constraint, we introduce electrical I/O efficiency, defined as the computational throughput achieved per electrical I/O channel. This metric complements conventional throughput-based evaluations by accounting for the interface overhead associated with large-scale optical computing systems. Figure 2f and Supplementary Table 1 compare our temporal matrix multiplication architecture with SOTA optical tensor multiplication approaches. Our approach achieves competitive matrix dimensions while requiring only a single optical I/O port, four electrical I/O connections and two active devices, resulting in substantially improved electrical I/O efficiency.

## Temporal diffractive neural networks with *in-situ* training

We extend the temporal computing platform from individual optical operators to multilayer TDNNs. The optical feedback loop enables forward propagation across multiple layers in the temporal domain without intermediate analogue-to-digital or digital-to-analogue conversion of temporal representations. In this architecture, IQ modulators are generalized from implementing specific optical transformations, e.g., a time lens for tensor processing, to programmable temporal masks that define trainable complex-valued weights.

Figure 3a illustrates the principle of our TDNNs, where the feedback architecture allows the entire multilayer network to be physically implemented using a single optical pathway. Taking handwritten digit recognition as an example, the input image is first flattened into a temporal vector and encoded onto the first layer through a temporal mask. Each temporal mask is represented by an arbitrary complex-valued waveform (Supplementary Fig. 10), which independently modulates the temporal neurons in each layer. During dispersive propagation, temporal neurons interact through the dispersive transfer function, establishing inter-layer connections analogous to diffraction in spatial diffractive neural networks. After multiple round trips through the optical feedback loop, the classification output is determined by the temporal position of the maximum response among ten output slots. As shown in Fig. 3d, the temporal evolution of intermediate representations and the final output can be directly monitored using a DSO. The TDNN achieves an inference accuracy of 95.3% on the Scikit-learn Digits dataset (Fig. 3e).

Beyond inference, the programmable temporal masks enable *in situ* training of the TDNNs. As illustrated in Fig. 3b, we develop a hardware–software co-design framework based on a digital twin [31,43,44]. The physical TDNN hardware, consisting of the optical transmitter, dispersive medium and receiver, performs forward propagation, while the software counterpart calculates gradients, performs backpropagation and updates the temporal masks based on the measured optical outputs. We validate this *in-situ* training capability using the Two Circles dataset (Supplementary Fig. 6). The training and testing accuracy evolution are shown in Fig. 3c, with the final test accuracy reaching 92%.

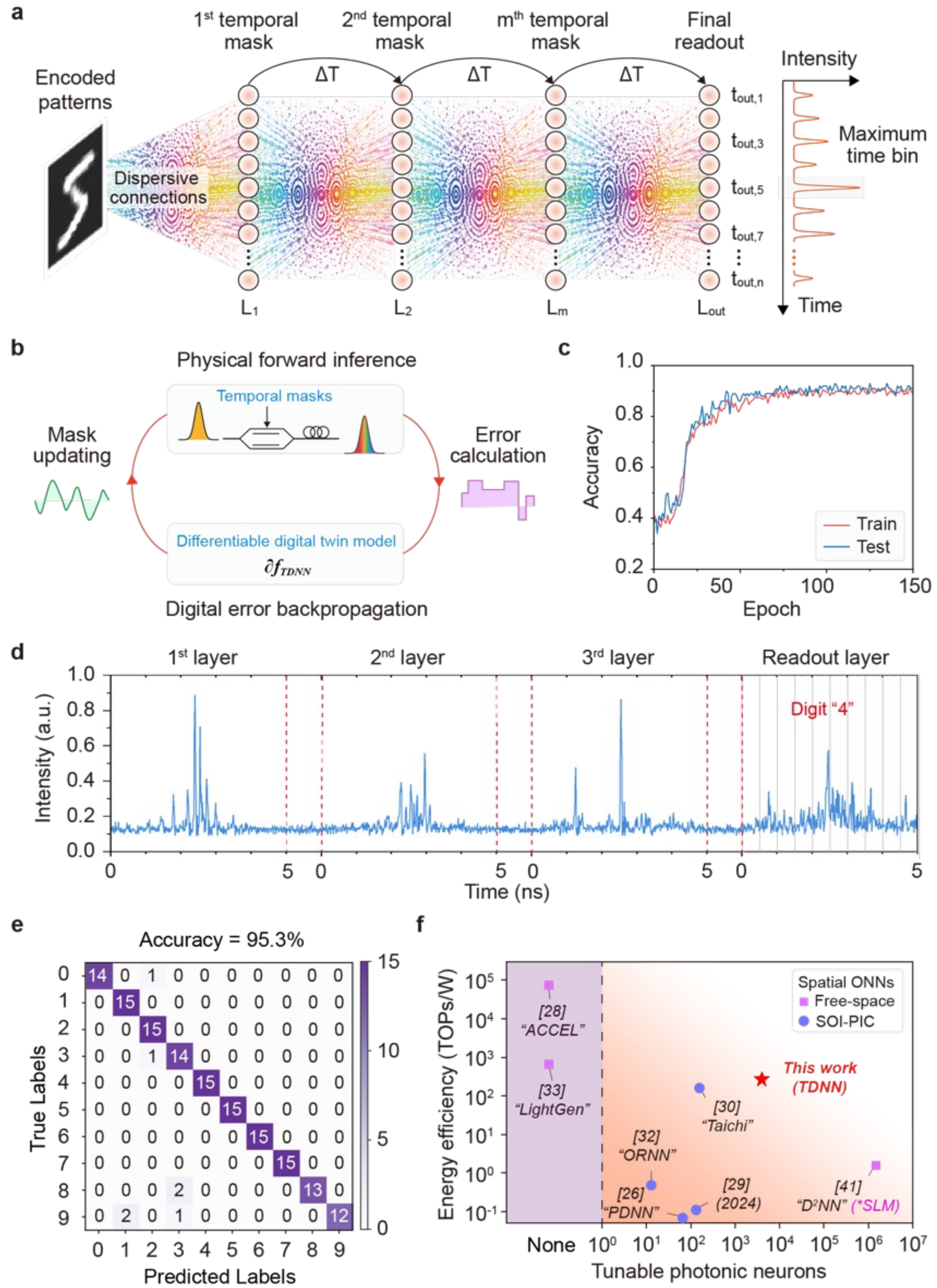

**Fig. 3. Temporal diffractive neural network. a,** The implementation of a TDNN on the digit handwriting images classification. After flattening the input image to a vector, the temporal optical signal passes through a temporal mask. The temporal masks are arbitrary complex-valued waveforms that modulate time neurons at different layers. Dispersion establishes connections between different time neurons between two layers. After multiple rounds of forward propagation through the optical loop, the modulator offers a window to read the output waveform at the final layer. **b,** Architecture of software-hardware twin networks for in-situ training. **c,** The train accuracy

and test accuracy of in-situ training for Two Circles dataset, and the final test-accuracy reaches 92%. **d,** Feature maps of each layers continuously shown in the digital storage oscilloscope. **e,** Confusion matrix of the Scikit-learn Digits Dataset classification. **f,** Comparison with SOTA ONNs.

These results demonstrate that TDNNs provide the essential attributes required for modern ONNs, including scalability, depth, and trainability. Figure 3f compares the energy efficiency and reconfigurability of TDNNs with SOTA spatial ONNs, showing an advantageous balance between computational scalability and hardware complexity. Furthermore, the temporal architecture enables flexible assignment of different optical operations across network layers. By combining tensor processing, diffractive propagation and time-lens-enabled Fourier transformation, TDNNs can implement diverse neural architectures, including fully connected, diffractive, convolutional and Fourier neural networks. Unlike conventional spatial ONNs that are often constrained by fixed physical layouts, temporal processing enables reconfiguration of network topology through programmable temporal operations. Therefore, the proposed TDNN constitutes a complete photonic computing paradigm by preserving the key advantages of spatial PNNs in a compact temporal platform.

## Temporal image and video generation

Motivated by the rapid progress of large language models and agent-based models, generative tasks have become a stringent benchmark for evaluating emerging photonic computing architectures. We therefore extend the TDNN beyond discriminative tasks and investigate its capability for accelerating generative models. As illustrated in Fig. 4a, the TDNN is integrated as the decoder of a variational autoencoder (VAE) framework for image and video generation. Temporal latent representations are directly processed by the TDNN, where each temporal sample acquired at a 100-GSa/s sampling rate corresponds to an image pixel. This direct temporal-to-spatial mapping enables high-frame-rate generation, including generation of 50×50 pixel images at frame rates exceeding 40 MHz.

The temporal latent representations can be generated either through electrically encoded signals or directly from physical temporal noise. Here, we demonstrate two complementary generation strategies enabled by the TDNN. First, we realize fully analog end-to-end random image generation by using amplified spontaneous emission (ASE) noise as the latent source (Fig. 4b). Conventional generative pipelines typically rely on digitally sampled Gaussian latent vectors followed by electrical modulation before optical processing[33]. In contrast, our approach directly harnesses the naturally available complex Gaussian statistics of optical noise in the temporal domain (Supplementary Fig. 7b), avoiding digital latent sampling and electronic encoding. Using

ASE noise as the input source, we generate MNIST images and compare the results with a conventional single-wavelength continuous-wave laser-based approach (Supplementary Fig. 8).

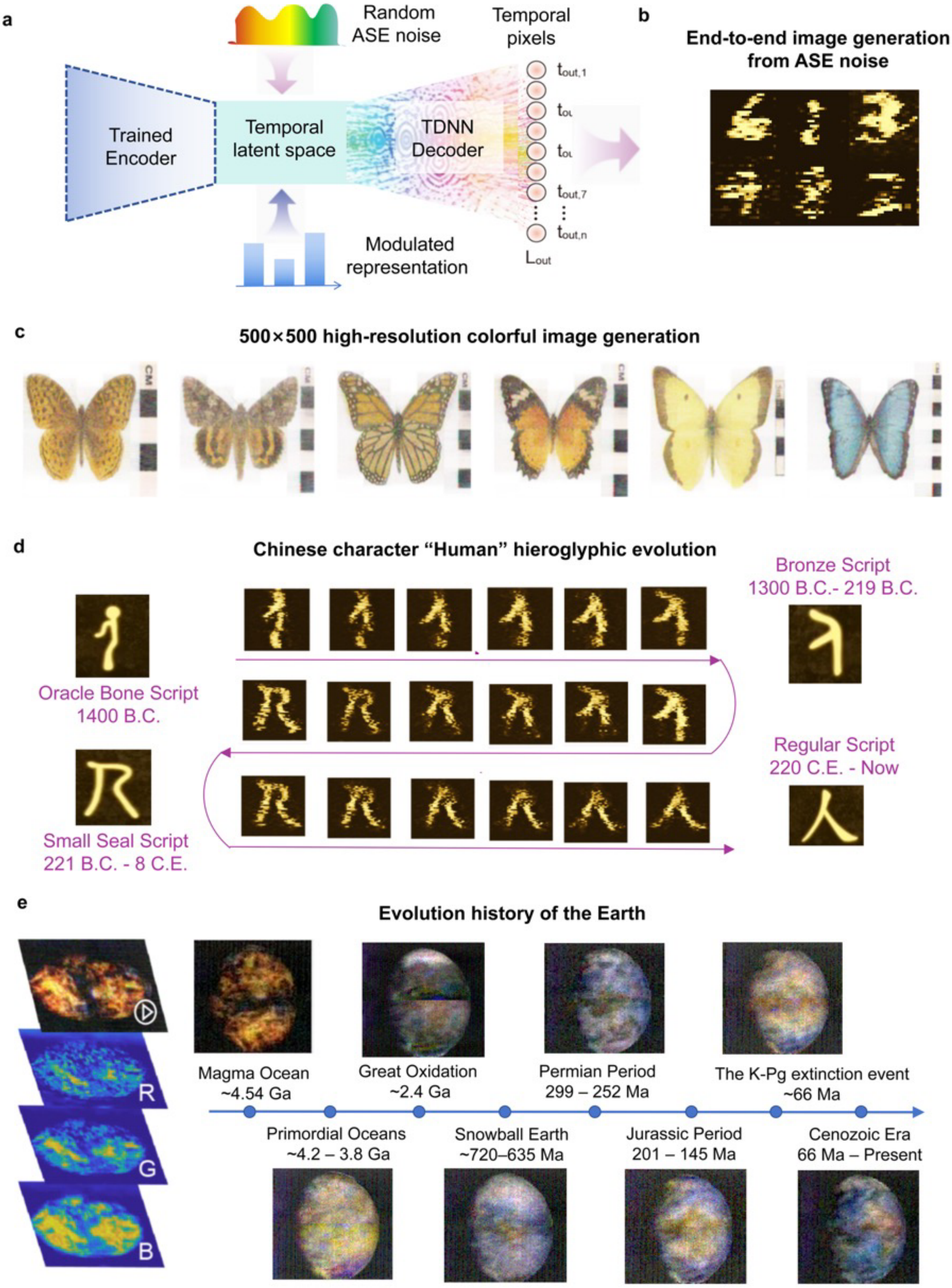

**Fig. 4. TDNN for generative tasks. a,** TDNN architecture for accelerating generative tasks, where TDNN acts as the decoder within a Variational Autoencoder (VAE) framework to generate images or videos at ultra-high frame rates. The temporal latent representation can be electrically modulated

or from natural temporal noise. **b,** End-to-end random MNIST image generation from amplified spontaneous emission (ASE) noise by TDNN. **c,** Experimental generation of high-resolution (500 ×500) colorful butterfly images, based on electrically modulated latent representations. Zoomed-in details are shown in Extended Data Fig. 1. **d,** Ultra-fast frame rate video generation by TDNN: the evolution of the Chinese character "Human", from oracle-bone script to regular script in grayscale. **e,** Experimental generation of RGB videos illustrating the evolutionary history of Earth.

To further demonstrate the scalability of the TDNN, we experimentally reconstructed and generated high-resolution 500×500 pixel colour images of butterflies (Fig. 4c). In contrast to the noise-driven MNIST demonstration, electrically encoded latent representations are employed in this experiment to provide precise control over complex image features. The reconstructed butterfly images resolve detailed textures and colour gradients, as illustrated in the enlarged views in Extended Data Fig. 1. The average peak signal-to-noise ratio (PSNR) and structural similarity index measure (SSIM) are 15.09 dB and 0.67, respectively. Additionally, the integration of the TDNN architecture with Stable Diffusion is utilized to generate random butterfly images from Gaussian distribution, which is described in Supplementary Fig. 9.

Beyond static image generation, we further extend TDNN-based generation towards sequential video generation by using a series of interpolated temporal latent representations. Feeding continuous latent trajectories into the TDNN enables visualization of the evolution of the Chinese character "Human" from Oracle-Bone Script to Regular Script in grayscale (Fig. 4d). Moreover, RGB video sequences depicting the evolutionary history of Earth are generated using the same temporal generative framework (Fig. 4e). Additionally, sampling from a Gaussian distribution enables the creation of random artistic renderings of celestial bodies (Supplementary Fig. 14), yielding a Fréchet Inception Distance (FID)[45] of 122 and an Inception Score (IS)[46] of 2.527 ± 0.121. These demonstrations highlight the versatility of TDNNs in handling complex, multi-frame temporal data.

## Discussion

Shifting the computational domain from space to time fundamentally changes the scaling paradigm for ONNs. Spatial ONNs are inherently constrained by their fixed physical topology upon fabrication, where increasing computational dimensions generally requires replication of optical components. This scaling strategy inevitably introduces additional electrical interfaces, packaging complexity, and system-level overhead, which can become limiting factors even when the optical core itself provides substantial computational throughput. Temporal ONNs[15,42,47-50], in contrast, reuse the same optical pathway across sequential temporal neurons, allowing network dimensions and connectivity to be encoded through time. Their scalability is therefore less governed by the number of physical components and more by available bandwidth, temporal aperture and characteristic system timescales.

The radio-frequency dimension has been extensively explored for photonic tensor processing, however, extracting parallel outputs encoded on different frequency channels typically requires frequency demultiplexing and additional detection hardware[17]. This readout complexity can partially offset the latency and energy advantages offered by optical processing. In contrast, our approach exploits temporal dispersive Fourier transformation enabled by a time-lens, which performs coherent optical summation and maps computational results onto resolvable temporal peaks. As a result, MMM can be executed on a single optical carrier and directly recovered in the time domain. Importantly, WDM can be introduced as an orthogonal scaling dimension, providing additional degrees of freedom for higher-order tensor operations. The ultimate matrix dimension is primarily determined by the modulation bandwidth, which defines the effective temporal numerical aperture and the available number of resolvable time-frequency modes. Increasing the bandwidth therefore offers a direct route towards larger computational size while maintaining the same optical architecture. Such scaling, however, requires precise synchronization between modulation, dispersion and electronic acquisition. As each computational element is transformed into a localized temporal peak, sampling away from the maximum introduces computational errors.

Our TDNN extends this concept beyond previously demonstrated time-wavelength-interleaved MVM accelerators[15,48] and time-synthetic ONNs[42,47] by enabling fully-programmable networks within a single reconfigurable optical path. Network depth is encoded into successive round trips, whereas neuron counts and complex-valued layer responses are independently defined through temporal masks. This separation between physical propagation and logical network configuration allows depth, width, and inter-neuron connectivity to be modified without redesigning the underlying optical hardware. Beyond fully connected layers, temporal Fourier transformation and spectral filtering can be incorporated into the same framework to enable convolutional processing. Furthermore, integrating semiconductor optical amplifiers into the feedback loop could extend the computational functionality by introducing all-optical nonlinear activation mechanisms[51].

Realizing this temporal architecture on an integrated platform will require addressing several remaining challenges. The present system relies on a long dispersive fiber to provide the group-delay spread required for temporal Fourier transformation and coupling between temporal neurons. Future implementations may exploit compact dispersive structures, including chirped Bragg gratings[52] or engineered photonic-crystal devices, to reduce the footprint while maintaining the required temporal aperture. In addition, accumulated loss during multiple round trips must be compensated for deeper networks. Potential integrated solutions include erbium-doped thin-film lithium niobate platforms[53] or heterogeneous integration of semiconductor gain medium[54].

In conclusion, we establish a photonic temporal computing paradigm, validating its completeness across multiple levels, spanning fundamental operator functions, the extension to trainable multilayer networks, and application-level models and tasks. By translating representative spatial optical computing primitives, including diffraction and tensor multiplication, into the temporal domain and further combining them with wavelength multiplexing and recurrent propagation, we demonstrate a unified platform supporting high-order tensor processing, trainable multilayer networks and generative models. The core contribution is that computational dimensionality becomes decoupled from physical hardware scaling. By transferring optical computation from spatial replication to temporal reuse, our framework provides a pathway towards scalable photonic intelligence while preserving the parallelism and dense connectivity that motivate optical neural networks. Furthermore, its compatibility with integrated platforms further suggests future opportunities for combining the computational advantages of free-space optical systems with the compactness and scalability of integrated photonics.

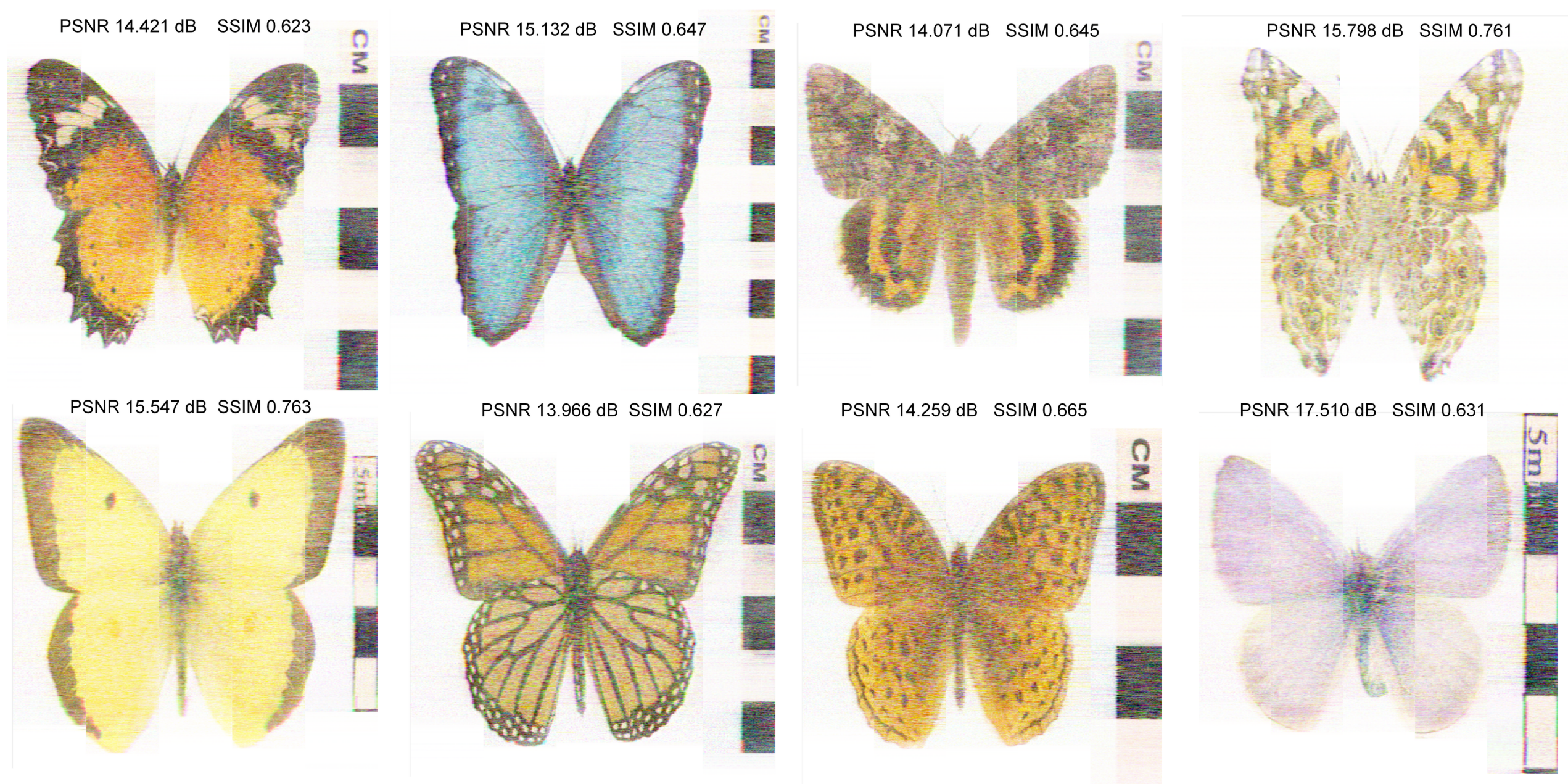

**Extended Data Fig. 1 | Experimental reconstruction of high-resolution (500×500) colorful butterfly images.**

**Acknowledgements**

X.H.G. acknowledges the support by National Key R&D Program of China (Grant No. 2023YFB2804702), Natural Science Foundation of China (NSFC) (Grant No. 62550072), Shanghai Science and Technology Innovation Action Plan (Grant No. 25LN3201000, 25JD1405500), Shanghai Municipal Science and Technology Major Project. Y.K.S acknowledges the support by Natural Science Foundation of China (NSFC) (Grant No. 62341508), Shanghai Science and Technology Innovation Action Plan (Grant No. 24JD1401500). J.L.X. acknowledges the support by Natural Science Foundation of China (NSFC) (Grant No. 62505175). Y.L.C acknowledges the support by Natural Science Foundation of China (NSFC) (Grant No. 625B2113).

**Author Contributions**

X.H.G., Z.P.S. initiated the project. Y.L.C., J.L.X., Y.F.Z. performed the calculation and simulation. X.H.G., Y.X.Z., Y.L.C., J.L.X., Y.M.H., Y.F.Z. designed the experiments. Y.C.Y., C.J.X., Z.S.L., X.W., J.W.Z., W.C., T. L., Y.X.Z., Q.B.Z assisted in the measurements. Y.L.C., J,L,X, Y.M.H., Y.X.Z. carried out the measurements. X.H.G., Y.L.C., J.L.X., Y.F.Z. analyzed the results and wrote the manuscript. X.H.G., and Z.P.S. supervised the project.

**Competing Interests**

The authors declare that they have no competing interests.

**Data Availability**

All raw data are available in the main text and the supplementary information.

**Code Availability**

The code will be made open access following acceptance.